%Paper: gr-qc/9208002
%From: "Andrea Pasquinucci" <pasquinu@puhep1.princeton.edu>
%Date: Tue, 11 Aug 92 15:10:54 EDT

%%%%%%%%%%%%%%%%%%%%%%%%%%%%%%%%%%%%%%%%%%%%%%%%%%%%%%%%%%%%%%%%%%%%%
%                          IASSNS-HEP-92/51, PUPT-1336, August 1992 %
%                                                                   %
%          Thermodynamics of Two-Dimensional Black-Holes            %
%             Chiara R. Nappi and Andrea Pasquinucci                %
%                                                                   %
%         This is a TeX file, all macros are already included       %
%%%%%%%%%%%%%%%%%%%%%%%%%%%%%%%%%%%%%%%%%%%%%%%%%%%%%%%%%%%%%%%%%%%%%
%
%                         " PANDA MACROS "
%
%   Double-page MUST be printed in LANDSCAPE orientation (or define
%    \Landspec).
%--------------------------------------------------------------------
%
%                     INTERACTIVE SECTION
%
%--------------------------------------------------------------------
%
\def\standardrisposta{s }\def\reducedrisposta{r }
\def\doublerisposta{d }\def\cartarisposta{e }\def\amsrisposta{y }
\newcount\ingrandimento \newcount\sinnota \newcount\dimnota
\newcount\unoduecol \newdimen\collhsize \newdimen\tothsize
\newdimen\fullhsize \newcount\controllorisposta \sinnota=1
\newskip\infralinea  \global\controllorisposta=0
\message{ ********    Welcome to PANDA macros (Plain TeX, AP, 1991)}
\message{ ******** }
\message{       You'll have to answer a few questions in lowercase.}
\message{>  Do you want it in double-page (d), reduced (r)}
\message{or standard format (s) ? }\read-1 to\risposta
\message{>  Do you want it in USA A4 (u) or EUROPEAN A4 (e)}
\message{paper size ? }\read-1 to\srisposta
\message{>  Do you have AMSFonts 2.0 (math) fonts (y/n) ? }
\read-1 to\arisposta
%
%--------------------------------------------------------------------
%
%             END INTERACTIVE SECTION - PAGE FORMATTING
%
%--------------------------------------------------------------------
%
%\def\risposta{d } \def\srisposta{u } \def\arisposta{n }
%
\ifx\risposta\standardrisposta \ingrandimento=1200
\message{>> This will come out UNREDUCED << }
\dimnota=2 \unoduecol=1 \global\controllorisposta=1 \fi
\ifx\risposta\reducedrisposta \ingrandimento=1095 \dimnota=1
\unoduecol=1  \global\controllorisposta=1
\message{>> This will come out REDUCED << } \fi
\ifx\risposta\doublerisposta \ingrandimento=1000 \dimnota=2
\unoduecol=2  \global\controllorisposta=1
\message{>> You must print this in LANDSCAPE orientation << } \fi
\ifnum\controllorisposta=0  \ingrandimento=1200
\message{>>> ERROR IN INPUT, I ASSUME STANDARD UNREDUCED FORMAT <<< }
\dimnota=2 \unoduecol=1 \fi
\magnification=\ingrandimento
%
%--------------------------------------------------------------------
%
%                        PARAMETERS SETTING
%
%  You can modify these parameters at your will (and resposability)
%--------------------------------------------------------------------
%
\newdimen\eucolumnsize \newdimen\eudoublehsize \newdimen\eudoublevsize
\newdimen\uscolumnsize \newdimen\usdoublehsize \newdimen\usdoublevsize
\newdimen\eusinglehsize \newdimen\eusinglevsize \newdimen\ussinglehsize
\newskip\standardbaselineskip \newdimen\ussinglevsize
\newskip\reducedbaselineskip \newskip\doublebaselineskip
\eucolumnsize=12.0truecm    % column h-size for european doublepage
                            % (12.0treucm default)
\eudoublehsize=25.5truecm   % sheet h-size for european duoblepage
                            % (25.5treucm default)
\eudoublevsize=6.5truein    % sheet v-size for european doublepage
                            % (6.5treuin default  or 17truecm?)
\uscolumnsize=4.4truein     % column h-size for american doublepage
                            % (4.4treuin default)
\usdoublehsize=9.4truein    % sheet h-size for american duoblepage
                            % (9.4treuin default)
\usdoublevsize=6.8truein    % sheet v-size for american doublepage
                            % (6.8treuin default)
\eusinglehsize=6.5truein    % sheet h-size for european singlepage
                            % (6.5truein default)
\eusinglevsize=24truecm     % sheet v-size for european singlepage
                            % (24truecm default)
\ussinglehsize=6.5truein    % sheet h-size for american singlepage
                            % (6.5truein default)
\ussinglevsize=8.9truein    % sheet v-size for american singlepage
                            % (8.9truein default)
\standardbaselineskip=16pt  % baselineskip for standard format
                            % (16pt default)
\reducedbaselineskip=14pt   % baselineskip for reduced format
                            % (14pt default)
\doublebaselineskip=12pt    % baselineskip for doublepage format
                            % (12pt default)
%
%  \Portoffset and \Landoffset define the horizontal and vertical
%  offsets respectively for portrait and landscape modes. Example:
%  \def\Portoffset{\voffset=.4truein\hoffset=.125truein}
%
\def\Portoffset{}
\def\Landoffset{}
%
%  \Landspec defines the \special command that sets the printer
%  to landscape mode without need to specify it directly in the
%  TeX to postscript translator (the command is site dependent).
%  Example: \def\Landspec{\special{ps: landscape}}
%
\def\Landspec{}
\tolerance=10000
\parskip 0pt plus 2pt  %\leftskip=0pt \rightskip=0pt
%
%   Do not modify anything of what follows
%                       (unless you know what you are doing!)
%--------------------------------------------------------------------
%
\ifx\risposta\standardrisposta \infralinea=\standardbaselineskip \fi
\ifx\risposta\reducedrisposta  \infralinea=\reducedbaselineskip \fi
\ifx\risposta\doublerisposta   \infralinea=\doublebaselineskip \fi
\ifnum\controllorisposta=0    \infralinea=\standardbaselineskip \fi
\ifx\risposta\doublerisposta   \Landoffset \else \Portoffset \fi
\ifx\risposta\doublerisposta \ifx\srisposta\cartarisposta
\tothsize=\eudoublehsize \collhsize=\eucolumnsize
\vsize=\eudoublevsize  \else  \tothsize=\usdoublehsize
\collhsize=\uscolumnsize \vsize=\usdoublevsize \fi \else
\ifx\srisposta\cartarisposta \tothsize=\eusinglehsize
\vsize=\eusinglevsize \else  \tothsize=\ussinglehsize
\vsize=\ussinglevsize \fi \collhsize=4.4truein \fi
%
%--------------------------------------------------------------------
%
%                            FONTS
%
%--------------------------------------------------------------------
%
\newcount\contaeuler \newcount\contacyrill \newcount\contaams
\font\ninerm=cmr9  \font\eightrm=cmr8  \font\sixrm=cmr6
\font\ninei=cmmi9  \font\eighti=cmmi8  \font\sixi=cmmi6
\font\ninesy=cmsy9  \font\eightsy=cmsy8  \font\sixsy=cmsy6
\font\ninebf=cmbx9  \font\eightbf=cmbx8  \font\sixbf=cmbx6
\font\ninett=cmtt9  \font\eighttt=cmtt8  \font\nineit=cmti9
\font\eightit=cmti8 \font\ninesl=cmsl9  \font\eightsl=cmsl8
\skewchar\ninei='177 \skewchar\eighti='177 \skewchar\sixi='177
\skewchar\ninesy='60 \skewchar\eightsy='60 \skewchar\sixsy='60
\hyphenchar\ninett=-1 \hyphenchar\eighttt=-1 \hyphenchar\tentt=-1
  \font\tencmmib=cmmib10  \newfam\cmmibfam
\skewchar\tencmmib='177   \font\tencmbsy=cmbsy10
\newfam\cmbsyfam  \skewchar\tencmbsy='60
\def\scaps{\cmcsc}  \font\tencmcsc=cmcsc10  \newfam\cmcscfam
\ifnum\ingrandimento=1095 
 
\font\bfone=cmbx10 at 10.95pt

\font\capsone=cmcsc10 at 10.95pt 

\else  
 
\font\bfone=cmbx10 at 12pt

\font\capsone=cmcsc10 at 12pt 
\fi
\def\chapterfont#1{\xdef\ttaarr{#1}}
\def\sectionfont#1{\xdef\ppaarr{#1}}
\def\ttaarr{\bf} \def\ppaarr{\sl}
%
%   AMS fonts (this works only if you have at least the 2.0
%              version of AMSFonts, otherwise say no)
%
\newfam\eufmfam \newfam\msamfam \newfam\msbmfam \newfam\eufbfam
\def\loadamsmath{\global\contaams=1 \ifx\arisposta\amsrisposta
\font\tenmsam=msam10 \font\ninemsam=msam9 \font\eightmsam=msam8
\font\sevenmsam=msam7 \font\sixmsam=msam6 \font\fivemsam=msam5
\font\tenmsbm=msbm10 \font\ninemsbm=msbm9 \font\eightmsbm=msbm8
\font\sevenmsbm=msbm7 \font\sixmsbm=msbm6 \font\fivemsbm=msbm5
\else \def\msbm{\bf} \fi \def\Bbb{\msbm} \def\symbl{\msam} \tenpoint}
\ifx\arisposta\amsrisposta
\font\sevenex=cmex7               %  reduced math symbols
\font\eightex=cmex8  \font\nineex=cmex9
\font\ninecmmib=cmmib9   \font\eightcmmib=cmmib8
\font\sevencmmib=cmmib7 \font\sixcmmib=cmmib6
\font\fivecmmib=cmmib5   \skewchar\ninecmmib='177
\skewchar\eightcmmib='177  \skewchar\sevencmmib='177
\skewchar\sixcmmib='177   \skewchar\fivecmmib='177
\font\ninecmbsy=cmbsy9    \font\eightcmbsy=cmbsy8
\font\sevencmbsy=cmbsy7  \font\sixcmbsy=cmbsy6
\font\fivecmbsy=cmbsy5   \skewchar\ninecmbsy='60
\skewchar\eightcmbsy='60  \skewchar\sevencmbsy='60
\skewchar\sixcmbsy='60    \skewchar\fivecmbsy='60
\font\ninecmcsc=cmcsc9    \font\eightcmcsc=cmcsc8     \else
\def\cmmib{\fam\cmmibfam\tencmmib}\textfont\cmmibfam=\tencmmib
\scriptfont\cmmibfam=\tencmmib \scriptscriptfont\cmmibfam=\tencmmib
\def\cmbsy{\fam\cmbsyfam\tencmbsy} \textfont\cmbsyfam=\tencmbsy
\scriptfont\cmbsyfam=\tencmbsy \scriptscriptfont\cmbsyfam=\tencmbsy
\scriptfont\cmcscfam=\tencmcsc \scriptscriptfont\cmcscfam=\tencmcsc
\def\cmcsc{\fam\cmcscfam\tencmcsc} \textfont\cmcscfam=\tencmcsc \fi
\catcode`@=11
\newskip\ttglue
\gdef\tenpoint{\def\rm{\fam0\tenrm}
  \textfont0=\tenrm \scriptfont0=\sevenrm \scriptscriptfont0=\fiverm
  \textfont1=\teni \scriptfont1=\seveni \scriptscriptfont1=\fivei
  \textfont2=\tensy \scriptfont2=\sevensy \scriptscriptfont2=\fivesy
  \textfont3=\tenex \scriptfont3=\tenex \scriptscriptfont3=\tenex
  \def\mcal{\fam2 \tensy}  \def\mmit{\fam1 \teni}
  \textfont\itfam=\tenit \def\it{\fam\itfam\tenit}
  \textfont\slfam=\tensl \def\sl{\fam\slfam\tensl}
  \textfont\ttfam=\tentt \scriptfont\ttfam=\eighttt
  \scriptscriptfont\ttfam=\eighttt  \def\tt{\fam\ttfam\tentt}
  \textfont\bffam=\tenbf \scriptfont\bffam=\sevenbf
  \scriptscriptfont\bffam=\fivebf \def\bf{\fam\bffam\tenbf}
     \ifx\arisposta\amsrisposta    \ifnum\contaams=1
  \textfont\msamfam=\tenmsam \scriptfont\msamfam=\sevenmsam
  \scriptscriptfont\msamfam=\fivemsam \def\msam{\fam\msamfam\tenmsam}
  \textfont\msbmfam=\tenmsbm \scriptfont\msbmfam=\sevenmsbm
  \scriptscriptfont\msbmfam=\fivemsbm \def\msbm{\fam\msbmfam\tenmsbm}
     \fi  \textfont3=\tenex \scriptfont3=\sevenex
  \scriptscriptfont3=\sevenex
  \def\cmmib{\fam\cmmibfam\tencmmib} \scriptfont\cmmibfam=\sevencmmib
  \textfont\cmmibfam=\tencmmib  \scriptscriptfont\cmmibfam=\fivecmmib
  \def\cmbsy{\fam\cmbsyfam\tencmbsy} \scriptfont\cmbsyfam=\sevencmbsy
  \textfont\cmbsyfam=\tencmbsy  \scriptscriptfont\cmbsyfam=\fivecmbsy
  \def\cmcsc{\fam\cmcscfam\tencmcsc} \scriptfont\cmcscfam=\eightcmcsc
  \textfont\cmcscfam=\tencmcsc \scriptscriptfont\cmcscfam=\eightcmcsc
     \fi            \tt \ttglue=.5em plus.25em minus.15em
  \normalbaselineskip=12pt
  \setbox\strutbox=\hbox{\vrule height8.5pt depth3.5pt width0pt}
  \let\sc=\eightrm \let\big=\tenbig   \normalbaselines
  \baselineskip=\infralinea  \rm}
\gdef\ninepoint{\def\rm{\fam0\ninerm}
  \textfont0=\ninerm \scriptfont0=\sixrm \scriptscriptfont0=\fiverm
  \textfont1=\ninei \scriptfont1=\sixi \scriptscriptfont1=\fivei
  \textfont2=\ninesy \scriptfont2=\sixsy \scriptscriptfont2=\fivesy
  \textfont3=\tenex \scriptfont3=\tenex \scriptscriptfont3=\tenex
  \def\mcal{\fam2 \ninesy}  \def\mmit{\fam1 \ninei}
  \textfont\itfam=\nineit \def\it{\fam\itfam\nineit}
  \textfont\slfam=\ninesl \def\sl{\fam\slfam\ninesl}
  \textfont\ttfam=\ninett \scriptfont\ttfam=\eighttt
  \scriptscriptfont\ttfam=\eighttt \def\tt{\fam\ttfam\ninett}
  \textfont\bffam=\ninebf \scriptfont\bffam=\sixbf
  \scriptscriptfont\bffam=\fivebf \def\bf{\fam\bffam\ninebf}
     \ifx\arisposta\amsrisposta  \ifnum\contaams=1
  \textfont\msamfam=\ninemsam \scriptfont\msamfam=\sixmsam
  \scriptscriptfont\msamfam=\fivemsam \def\msam{\fam\msamfam\ninemsam}
  \textfont\msbmfam=\ninemsbm \scriptfont\msbmfam=\sixmsbm
  \scriptscriptfont\msbmfam=\fivemsbm \def\msbm{\fam\msbmfam\ninemsbm}
     \fi  \textfont3=\nineex \scriptfont3=\sevenex
  \scriptscriptfont3=\sevenex
  \def\cmmib{\fam\cmmibfam\ninecmmib}  \textfont\cmmibfam=\ninecmmib
  \scriptfont\cmmibfam=\sixcmmib \scriptscriptfont\cmmibfam=\fivecmmib
  \def\cmbsy{\fam\cmbsyfam\ninecmbsy}  \textfont\cmbsyfam=\ninecmbsy
  \scriptfont\cmbsyfam=\sixcmbsy \scriptscriptfont\cmbsyfam=\fivecmbsy
  \def\cmcsc{\fam\cmcscfam\ninecmcsc} \scriptfont\cmcscfam=\eightcmcsc
  \textfont\cmcscfam=\ninecmcsc \scriptscriptfont\cmcscfam=\eightcmcsc
     \fi            \tt \ttglue=.5em plus.25em minus.15em
  \normalbaselineskip=11pt
  \setbox\strutbox=\hbox{\vrule height8pt depth3pt width0pt}
  \let\sc=\sevenrm \let\big=\ninebig \normalbaselines\rm}
\gdef\eightpoint{\def\rm{\fam0\eightrm}
  \textfont0=\eightrm \scriptfont0=\sixrm \scriptscriptfont0=\fiverm
  \textfont1=\eighti \scriptfont1=\sixi \scriptscriptfont1=\fivei
  \textfont2=\eightsy \scriptfont2=\sixsy \scriptscriptfont2=\fivesy
  \textfont3=\tenex \scriptfont3=\tenex \scriptscriptfont3=\tenex
  \def\mcal{\fam2 \eightsy}  \def\mmit{\fam1 \eighti}
  \textfont\itfam=\eightit \def\it{\fam\itfam\eightit}
  \textfont\slfam=\eightsl \def\sl{\fam\slfam\eightsl}
  \textfont\ttfam=\eighttt \scriptfont\ttfam=\eighttt
  \scriptscriptfont\ttfam=\eighttt \def\tt{\fam\ttfam\eighttt}
  \textfont\bffam=\eightbf \scriptfont\bffam=\sixbf
  \scriptscriptfont\bffam=\fivebf \def\bf{\fam\bffam\eightbf}
     \ifx\arisposta\amsrisposta   \ifnum\contaams=1
  \textfont\msamfam=\eightmsam \scriptfont\msamfam=\sixmsam
  \scriptscriptfont\msamfam=\fivemsam \def\msam{\fam\msamfam\eightmsam}
  \textfont\msbmfam=\eightmsbm \scriptfont\msbmfam=\sixmsbm
  \scriptscriptfont\msbmfam=\fivemsbm \def\msbm{\fam\msbmfam\eightmsbm}
     \fi  \textfont3=\eightex \scriptfont3=\sevenex
  \scriptscriptfont3=\sevenex
  \def\cmmib{\fam\cmmibfam\eightcmmib}  \textfont\cmmibfam=\eightcmmib
  \scriptfont\cmmibfam=\sixcmmib \scriptscriptfont\cmmibfam=\fivecmmib
  \def\cmbsy{\fam\cmbsyfam\eightcmbsy}  \textfont\cmbsyfam=\eightcmbsy
  \scriptfont\cmbsyfam=\sixcmbsy \scriptscriptfont\cmbsyfam=\fivecmbsy
  \def\cmcsc{\fam\cmcscfam\eightcmcsc} \scriptfont\cmcscfam=\eightcmcsc
  \textfont\cmcscfam=\eightcmcsc \scriptscriptfont\cmcscfam=\eightcmcsc
     \fi             \tt \ttglue=.5em plus.25em minus.15em
  \normalbaselineskip=9pt
  \setbox\strutbox=\hbox{\vrule height7pt depth2pt width0pt}
  \let\sc=\sixrm \let\big=\eightbig \normalbaselines\rm}
\gdef\tenbig#1{{\hbox{$\left#1\vbox to8.5pt{}\right.\n@space$}}}
\gdef\ninebig#1{{\hbox{$\textfont0=\tenrm\textfont2=\tensy
   \left#1\vbox to7.25pt{}\right.\n@space$}}}
\gdef\eightbig#1{{\hbox{$\textfont0=\ninerm\textfont2=\ninesy
   \left#1\vbox to6.5pt{}\right.\n@space$}}}
 %for 10-pt math in 9-pt territory
\def\alternativefont#1#2{\ifx\arisposta\amsrisposta \relax \else
\xdef#1{#2} \fi}
\global\contaeuler=0 \global\contacyrill=0 \global\contaams=0
%
%--------------------------------------------------------------------
%
%                            MACROS
%
%--------------------------------------------------------------------
%
\newbox\fotlinebb \newbox\hedlinebb \newbox\leftcolumn
\gdef\makeheadline{\vbox to 0pt{\vskip-22.5pt
     \fullline{\vbox to8.5pt{}\the\headline}\vss}\nointerlineskip}
\gdef\makehedlinebb{\vbox to 0pt{\vskip-22.5pt
     \fullline{\vbox to8.5pt{}\copy\hedlinebb\hfil
     \line{\hfill\the\headline\hfill}}\vss} \nointerlineskip}
\gdef\makefootline{\baselineskip=24pt \fullline{\the\footline}}
\gdef\makefotlinebb{\baselineskip=24pt
    \fullline{\copy\fotlinebb\hfil\line{\hfill\the\footline\hfill}}}
\gdef\doubleformat{\shipout\vbox{\Landspec\makehedlinebb
     \fullline{\box\leftcolumn\hfil\columnbox}\makefotlinebb}
     \advancepageno}
\gdef\columnbox{\leftline{\pagebody}}
\gdef\line#1{\hbox to\hsize{\hskip\leftskip#1\hskip\rightskip}}
\gdef\fullline#1{\hbox to\fullhsize{\hskip\leftskip{#1}%
\hskip\rightskip}}
\gdef\footnote#1{\let\@sf=\empty
         \ifhmode\edef\#sf{\spacefactor=\the\spacefactor}\/\fi
         #1\@sf\vfootnote{#1}}
\gdef\vfootnote#1{\insert\footins\bgroup
         \ifnum\dimnota=1  \eightpoint\fi
         \ifnum\dimnota=2  \ninepoint\fi
         \ifnum\dimnota=0  \tenpoint\fi
         \interlinepenalty=\interfootnotelinepenalty
         \splittopskip=\ht\strutbox
         \splitmaxdepth=\dp\strutbox \floatingpenalty=20000
         \leftskip=\oldssposta \rightskip=\olddsposta
         \spaceskip=0pt \xspaceskip=0pt
         \ifnum\sinnota=0   \textindent{#1}\fi
         \ifnum\sinnota=1   \item{#1}\fi
         \footstrut\futurelet\next\fo@t}
\gdef\fo@t{\ifcat\bgroup\noexpand\next \let\next\f@@t
             \else\let\next\f@t\fi \next}
\gdef\f@@t{\bgroup\aftergroup\@foot\let\next}
\gdef\f@t#1{#1\@foot} \gdef\@foot{\strut\egroup}
\gdef\footstrut{\vbox to\splittopskip{}}
\skip\footins=\bigskipamount
\count\footins=1000  \dimen\footins=8in
\catcode`@=12
\tenpoint
\ifnum\unoduecol=1 \hsize=\tothsize   \fullhsize=\tothsize \fi
\ifnum\unoduecol=2 \hsize=\collhsize  \fullhsize=\tothsize \fi
\global\let\lrcol=L
\ifnum\unoduecol=1 \output{\plainoutput{\ifnum\tipbnota=2
\clearnmbnota\fi}} \fi
\ifnum\unoduecol=2 \output{\if L\lrcol
     \global\setbox\leftcolumn=\columnbox
     \global\setbox\fotlinebb=\line{\hfill\the\footline\hfill}
     \global\setbox\hedlinebb=\line{\hfill\the\headline\hfill}
     \advancepageno  \global\let\lrcol=R
     \else  \doubleformat \global\let\lrcol=L \fi
     \ifnum\outputpenalty>-20000 \else\dosupereject\fi
     \ifnum\tipbnota=2\clearnmbnota\fi }\fi
\def\ifdoublepage{\ifnum\unoduecol=2 }
\gdef\yespagenumbers{\footline={\hss\tenrm\folio\hss}}
\gdef\ciao{\par\vfill\supereject \ifnum\unoduecol=2
     \if R\lrcol  \headline={}\nopagenumbers\null\vfill\eject
     \fi\fi \end}
\newskip\olddsposta \newskip\oldssposta
\global\oldssposta=\leftskip \global\olddsposta=\rightskip
 \def\newline{\hfil\break}
\def\jump{\vskip\baselineskip} \newskip\iinnffrr
\def\sjump{\iinnffrr=\baselineskip
          \divide\iinnffrr by 2 \vskip\iinnffrr}
\def\bjump{\vskip\baselineskip \vskip\baselineskip}
\newcount\nmbnota  \def\clearnmbnota{\global\nmbnota=0}
\newcount\tipbnota 

\def\note#1{\global\advance\nmbnota by 1 \ifnum\tipbnota=1
    \footnote{$^{\rm\nttlett}$}{#1} \else {\ifnum\tipbnota=2
    \footnote{$^{\nttsymb}$}{#1}
    \else\footnote{$^{\the\nmbnota}$}{#1}\fi}\fi}
\def\nttlett{\ifcase\nmbnota \or a\or b\or c\or d\or e\or f\or
g\or h\or i\or j\or k\or l\or m\or n\or o\or p\or q\or r\or
s\or t\or u\or v\or w\or y\or x\or z\fi}
\def\nttsymb{\ifcase\nmbnota \or\dag\or\sharp\or\ddag\or\star\or
\natural\or\flat\or\clubsuit\or\diamondsuit\or\heartsuit
\or\spadesuit\fi}   \clearnmbnota
\def\numberfootnote{\global\tipbnota=0} \numberfootnote
\def\setnote#1{\expandafter\xdef\csname#1\endcsname{
\ifnum\tipbnota=1 {\rm\nttlett} \else {\ifnum\tipbnota=2
{\nttsymb} \else \the\nmbnota\fi}\fi} }
\def\formula{$$} \def\endformula{\eqno\numero $$}
\def\fr{\formula} \def\efr{\endformula}
\newcount\frmcount \def\clearfrmcount{\global\frmcount=0}
\def\numero{\global\advance\frmcount by 1   \ifnum\indappcount=0
  {\ifnum\cpcount <1 {\hbox{\rm (\the\frmcount )}}  \else
  {\hbox{\rm (\the\cpcount .\the\frmcount )}} \fi}  \else
  {\hbox{\rm (\applett .\the\frmcount )}} \fi}
\def\nameformula#1{\global\advance\frmcount by 1%
\ifnum\draftnum=0  {\ifnum\indappcount=0%
{\ifnum\cpcount<1\xdef\spzzttrra{(\the\frmcount )}%
\else\xdef\spzzttrra{(\the\cpcount .\the\frmcount )}\fi}%
\else\xdef\spzzttrra{(\applett .\the\frmcount )}\fi}%
\else\xdef\spzzttrra{(#1)}\fi%
\expandafter\xdef\csname#1\endcsname{\spzzttrra}
\eqno \ifnum\draftnum=0 {\ifnum\indappcount=0
  {\ifnum\cpcount <1 {\hbox{\rm (\the\frmcount )}}  \else
  {\hbox{\rm (\the\cpcount .\the\frmcount )}}\fi}   \else
  {\hbox{\rm (\applett .\the\frmcount )}} \fi} \else (#1) \fi $$}
\def\nfr{\nameformula}    \def\numali{\numero}
\def\nameali#1{\global\advance\frmcount by 1%
\ifnum\draftnum=0  {\ifnum\indappcount=0%
{\ifnum\cpcount<1\xdef\spzzttrra{(\the\frmcount )}%
\else\xdef\spzzttrra{(\the\cpcount .\the\frmcount )}\fi}%
\else\xdef\spzzttrra{(\applett .\the\frmcount )}\fi}%
\else\xdef\spzzttrra{(#1)}\fi%
\expandafter\xdef\csname#1\endcsname{\spzzttrra}
  \ifnum\draftnum=0  {\ifnum\indappcount=0
  {\ifnum\cpcount <1 {\hbox{\rm (\the\frmcount )}}  \else
  {\hbox{\rm (\the\cpcount .\the\frmcount )}}\fi}   \else
  {\hbox{\rm (\applett .\the\frmcount )}} \fi} \else (#1) \fi}
\clearfrmcount
\newcount\cpcount \def\clearcpcount{\global\cpcount=0}
\newcount\subcpcount \def\clearsubcpcount{\global\subcpcount=0}
\newcount\appcount \def\clearappcount{\global\appcount=0}
\newcount\indappcount \def\clearindappcount{\indappcount=0}
\newcount\sottoparcount 
 \newcount\draftnum \clearappcount
\clearindappcount \global\draftnum=0
\newcount\connttrre  \def\clearconnttrre{\global\connttrre=0}
\newcount\countref  \def\clearcountref{\global\countref=0}
\clearcountref
\def\chapter#1{\global\advance\cpcount by 1 \clearfrmcount
                 \goodbreak\null\vbox{\jump\nobreak
                 \clearsubcpcount\clearindappcount
                 \itemitem{\ttaarr\the\cpcount .\qquad}{\ttaarr #1}
                 \par\nobreak\jump\sjump}\nobreak}
\def\section#1{\global\advance\subcpcount by 1 \goodbreak\null
               \vbox{\sjump\nobreak\ifnum\indappcount=0
                 {\ifnum\cpcount=0 {\itemitem{\ppaarr
               .\the\subcpcount\quad\enskip\ }{\ppaarr #1}\par} \else
                 {\itemitem{\ppaarr\the\cpcount .\the\subcpcount\quad
                  \enskip\ }{\ppaarr #1} \par}  \fi}
                \else{\itemitem{\ppaarr\applett .\the\subcpcount\quad
                 \enskip\ }{\ppaarr #1}\par}\fi\nobreak\jump}\nobreak}
\clearsubcpcount
\clearappcount \clearindappcount
\def\references{\goodbreak\null\vbox{\jump\nobreak
   \itemitem{}{\ttaarr References} \nobreak\jump\sjump}\nobreak}
\def\introduction{\clearindappcount\clearappcount\clearcpcount
                  \clearsubcpcount\goodbreak\null\vbox{\jump\nobreak
  \itemitem{}{\ttaarr Introduction} \nobreak\jump\sjump}\nobreak}
\clearcpcount\clearcountref
\def\acknowledgements{\goodbreak\null\vbox{\jump\nobreak
\itemitem{ }{\ttaarr Acknowledgements} \nobreak\jump\sjump}\nobreak}
%
%     Maximum number of references = 200
%     boxes 50 -> 250 reserved for references
%
\catcode`@=11
\gdef\Ref#1{\expandafter\ifx\csname @rrxx@#1\endcsname\relax%
{\global\advance\countref by 1%
\ifnum\countref>200%
\message{>>> ERROR: maximum number of references exceeded <<<}%
\expandafter\xdef\csname @rrxx@#1\endcsname{0}\else%
\expandafter\xdef\csname @rrxx@#1\endcsname{\the\countref}\fi}\fi%
\ifnum\draftnum=0 \csname @rrxx@#1\endcsname \else#1\fi}
\gdef\beginref{\ifnum\draftnum=0  \gdef\Rref{\fairef}
\gdef\endref{\scriviref} \else\relax\fi \parskip 2pt plus 2pt
\baselineskip=12pt}
\def\Reflab#1{[#1]} \gdef\Rref#1#2{\item{\Reflab{#1}}{#2}}
\gdef\endref{\relax}  \newcount\conttemp
\gdef\fairef#1#2{\expandafter\ifx\csname @rrxx@#1\endcsname\relax
{\global\conttemp=0
\message{>>> ERROR: reference [#1] not defined <<<} } \else
{\global\conttemp=\csname @rrxx@#1\endcsname } \fi
\global\advance\conttemp by 50
\global\setbox\conttemp=\hbox{#2} }
\gdef\scriviref{\clearconnttrre\conttemp=50
\loop\ifnum\connttrre<\countref \advance\conttemp by 1
\advance\connttrre by 1
\item{\Reflab{\the\connttrre}}{\unhcopy\conttemp} \repeat}
\clearcountref \clearconnttrre
\catcode`@=12
\ifx\oldchi\undefined \let\oldchi=\chi
  \def\cchi{{\raise 1pt\hbox{$\oldchi$}}} \let\chi=\cchi \fi
  
\def\del{\partial}   
\def\frac#1#2{{\textstyle{#1 \over #2}}}
\def\noblackbox{\overfullrule=0pt}
\def\yesblackbox{\overfullrule=5pt}
\null
%
%--------------------------------------------------------------------
%
%                       THE TEXT FOLLOWS
%
%--------------------------------------------------------------------
%
\loadamsmath \chapterfont{\bfone} \sectionfont{\scaps}
\def\scstyle{\scriptstyle}
\nopagenumbers
{\baselineskip=12pt
\line{\hfill IASSNS-HEP-92/51}
\line{\hfill PUPT-1336}
\line{\hfill gr-qc/9208002}
\line{\hfill August, 1992}}
{\baselineskip=14pt
\ifdoublepage \bjump\bjump\else\vfill\fi
\centerline{\capsone Thermodynamics of Two-Dimensional Black-Holes}
\bjump\bjump
\centerline{\scaps Chiara R. Nappi}
\sjump
\centerline{\sl School of Natural Science, Institute for Advanced Study,}
\centerline{\sl Olden Lane, Princeton, NJ 08540, USA}
\jump
\centerline{and}
\jump
\centerline{\scaps Andrea Pasquinucci}
\sjump
\centerline{\sl Joseph Henry Laboratories, Department of Physics,}
\centerline{\sl Princeton University, Princeton, NJ 08544, USA}
\vfill
\ifdoublepage \eject\null\vfill\else\bjump\fi
\centerline{\capsone ABSTRACT}
\sjump
\noindent We explore the thermodynamics of a general class of two
dimensional dilatonic black-holes. A simple prescription is given that
allows us to compute the mass, entropy and thermodynamic
potentials, with results in agreement with those obtained by other
methods, when available.

\ifdoublepage\vfill\else\bjump\bjump\jump\fi
\pageno=0 \eject }
\yespagenumbers\pageno=1
%\doublespacing
%
\introduction
The original paper [\Ref{Witten}] on two dimensional black-holes
in string theory started a renewed interest in the longstanding
problem of formulating a non trivial theory of two dimensional
gravity. Progress has been made in finding solutions of WZW
models, which are exactly conformal field theories, as well as in
finding solutions of the low energy string effective actions.
Here we will restrict our analysis to the latter kind of models,
namely to effective actions of dilaton gravity in two dimensions.
A vast class of these models is classically soluble,
even in the presence of a potential for the dilaton field. For
instance, for a dilaton potential of the type produced by string
loops, the solutions have very interesting spacetime geometries,
more complex than the Reissner-Nordstr{\o}m one. They generically
exhibit more than two horizons, and the black-hole tiles the entire
plane in the Penrose diagram [\Ref{YN}].
Some of these models have proved useful as toy models to investigate
the evaporation and formation of black-holes [\Ref{Callan}], as well
as for more formal pursuits such as the study of perturbative
renormalizability [\Ref{Odin}].

These models also appear to be interesting laboratories for questions
about black-hole thermodynamics. We will use them as such, and find a
simple prescription to compute all thermodynamic functions.
We will write the on-shell action as the sum of two boundary terms.
One of them, related to the second fundamental form of  the
boundary, will give the ({\scaps adm})
mass   when computed at infinity and the entropy
when computed on the horizon. The other
boundary term vanishes at infinity and gives the charges and chemical
potentials when evaluated at the horizon. The sum of these two terms
will reproduce the free energy.

This prescription has similarities with methods advocated
in the context of higher dimensional black-holes
[\Ref{GH},\Ref{York},\Ref{Teit},\Ref{Wilc}].
As analogous ones, it is rather heuristic
and it is an interesting challenge to find a fully satisfying
theoretical justification.
\chapter{The model and its solutions}
Our action is
\fr
I\ =\ \int d^2x \sqrt{-g}\, e^{-2\phi} \left[ R +\gamma
g^{ab}\del_a\phi\del_b\phi -\frac14 e^{\epsilon\phi} F^2 +
V(\phi)\right]\ .
\nfr{action}
Particular cases of eq. \action\  reduce to a number
of well known models of two dimensional gravity [\Ref{Jactei}], as well
as to the string bosonic effective action, and to the heterotic
string effective action [\Ref{YN}].

The model is also connected with four-dimensional spherically
symmetric gravity. Indeed from four-dimensional pure gravity
$$
I^{(4)} \ =\ \int d^4 x \, \sqrt{-g^{(4)}} \, R^{(4)}
\efr
by using  spherical coordinates
$$
^{(4)}ds^2\ =\  ^{(2)}ds^2 +  e^{-2\phi}\Omega^2
\efr
one gets
$$
I^{(4)}= \int d^2x \sqrt{-g^{(2)}}\left[e^{-2\phi}(R^{(2)} +2\nabla^a
\phi\nabla_a\phi + 2e^{2\phi}) + 4\nabla^a(e^{-2\phi}\nabla_a\phi)
\right] \nfr{four}
which, apart from a total derivative usually neglected, is of the form
of eq. \action\ with $\gamma =2$ and an appropriate potential for the
dilaton.

Our interest will focus on static configurations. We will work in the
gauge
$$
ds^2\ =\ - g(x) dt^2 + {1\over g(x)} dx^2\ .
\nfr{gauge}
The equations of motion are then
$$\eqalignno{
(g\phi')' -2g\phi'^2 -\frac14f^2e^{\epsilon\phi} + \frac12 V(\phi)
&\  =\ 0 \cr  2\phi'' + (\gamma - 4)\phi'^2 &\ =\ 0\cr
(fe^{(\epsilon -2)\phi})'&\ =\ 0 &\nameali{equations}\cr}
$$
where $^\prime = \del /\del x$ and $F = f(x)\, dx \wedge dt$.
Substituting eqs. \equations\ into the action eq. \action\
(without using any explicit solution), we get
$$
I_{\,{\displaystyle\vert}_{\scstyle\rm on\ shell}}\ =\ -\int dt dx \,
\del_x \left[ 4 e^{-2\phi} g \del_x \phi + e^{-2\phi} \del_x g
\right]_{\,{\displaystyle\vert}_{\scstyle\rm on\ shell}}\ .
\nfr{shell}
This relation, already mentioned for $\gamma=4$ and no dilaton
potential in ref. [\Ref{GP}], is actually valid for all the models
in eq. \action. It is the generalization of an analogous relation
for 4D pure gravity. Since on shell $R^{(4)}=0$, it follows from eq.
\four\ that
$$\eqalignno{
 \int d^2x \sqrt{-g^{(2)}}\,e^{-2\phi}&\left[R^{(2)} +\,  2\nabla^a\phi
\nabla_a\phi + \, 2e^{2\phi}
\right]_{\,{\displaystyle\vert}_{\scstyle\rm on\ shell}}
&\nameali{nshell}
\cr &\quad=\ - \int d^2x\sqrt{-g^{(2)}}\, \nabla^a\left[4e^{-2\phi}
\nabla_a\phi\right]_{\,{\displaystyle\vert}_{\scstyle\rm on\ shell}}
\ .\cr} $$
Indeed eq. \nshell\ is equivalent to eq. \shell\ since for 4D pure
gravity $\int d^2x \, \partial_x(e^{-2\phi}\partial_xg)=0$. Eq.
\shell\ can be written in fully covariant form in terms of the boundary
term in eq. \nshell\ and the second fundamental form of the boundary in
the 2D metric.

Another way of understanding the result in eq. \shell\ can be achieved
if one computes to the leading order the variation of eq. \action\
under the dilaton transformation $\phi\rightarrow \phi + a$, where $a$
is space dependent and with compact support. The above transformation
is not a symmetry of eq. \action, not even in the absence of potential;
however, on shell, where the variation  is zero anyway, one gets
$$
I_{\,{\displaystyle\vert}_{\scstyle\rm on\ shell}}\ =\  \int d^2x
\left[-\gamma\, \nabla^a(e^{-2\phi}\nabla_a\phi) + \frac12 e^{-2\phi}
{{\partial V} \over {\partial \phi}}\right]
\nfr{symm}
By using the equation of motions in their covariant form [\Ref{LN}],
one can rewrite eq. \symm\ in the more illuminating form
\noblackbox
$$
I_{\,{\displaystyle\vert}_{\scstyle\rm on\ shell}} =  \int d^2x
\left[- 4\nabla^a(e^{-2\phi}\nabla_a\phi) + e^{-2\phi}\left(R +
2\nabla^2\phi + (\gamma -4)(\nabla\phi)^2\right)\right]
\nfr{newsymm}
Finally, using eqs. \equations, one can show that also the second
term in eq. \newsymm\ is a total derivative independent of $\gamma$,
and is exactly the second term in eq. \shell.  \yesblackbox

The equation \shell\ will be the main input in our derivation of the
thermodynamics relations, which will basically consist in evaluating
eq. \shell\ on specific solutions of the equations of motion.
To that purpose we list here the solutions of eqs. \equations\ we will
consider.
\sjump
\noindent {\sl Solution 1 [\Ref{YN}]}\newline
Setting
$$\eqalignno{ & \gamma\ =\ 4\ , \qquad F\ =\ f(x)\, dx \wedge dt \ ,
\qquad V(\phi)\ =\ \sum_{n\geq 0} a_n e^{2n\phi(x)} \cr
& \epsilon\ =\ 0\ ,\qquad a_0\ =\ Q^2 \ ,\qquad a_1\ =\ 0&\numali\cr}
$$
one gets
$$\eqalignno{ & \phi(x)\ =\ \phi_0 - \frac12 Q x\ , \qquad
g(x)\ =\ 1 - 2me^{-Qx} + \sum_{n\geq 2} {b_n\over (1-n)Q^2} e^{-nQx}\cr
& f(x)\ =\ f_0 e^{2\phi(x)} &\nameali{solone}\cr}
$$
where $ b_n = e^{2n\phi_0} ( a_n -\frac12 \delta_{n,2} f_0^2)$
and it is convenient to set $ q^2 = f_0^2 e^{4\phi_0}/2Q^2$.

A particular solution is given by $b_2=-q^2Q^2$, $b_n=0$ ($n > 2$),
$a_n=0$ ($ n\geq 1$), this is the charged heterotic black-hole of ref.
[\Ref{YN}]. The above potential might be generated by string loops
corrections. As discussed in ref. [\Ref{YN}] a non-trivial potential
of this kind generically gives rise to interesting space-time
geometries with multiple horizons.
\sjump
\noindent {\sl Solution 2 [\Ref{LN}]}\newline
Setting
$$\eqalignno{ & \gamma\ =\ 2\ , \qquad F\ =\ f(x)\, dx \wedge dt
\ ,\qquad V(\phi)\ =\ \sum_{n\geq 0} a_n e^{2n\phi(x)}\cr
&\epsilon\ =\ 0\ ,\qquad a_0\ =\ 0\ ,\qquad a_1\ =\ 2 &\numali\cr}
$$
one gets
$$\eqalignno{ & \phi(x)\ =\ -\log (x)\ , \qquad
g(x)\ =\ 1 - {2m\over x} + \sum_{n\geq 2} {b_n\over 6-4n} x^{2-2n}\cr
& f(x)\ =\ f_0 e^{2\phi(x)} &\nameali{soltwo}\cr}
$$
where $b_n=( a_n -\frac12 \delta_{n,2}
f_0^2)$ and it is convenient to set $q^2 = \frac14 f_0^2$ .

If all the $b_n$ are zero except $b_1$, then of course one gets the
Schwarzschild black-hole. If $b_2=-2q^2$, $b_n=0$ ($n\geq 3$), $a_n=0$
($n\neq 1$), one gets the charged Reissner-Nordstr{\o}m  black-hole.
\sjump
\noindent{\sl Solutions with $0<\gamma<4$ }\newline
This is a new class of solutions of the equations of motion eqs.
\equations\ for $0<\gamma<4$ of which Solution 2 is the one with
$\gamma =2 $. Let
$$\eqalignno{& 0 <\gamma < 4\ ,\qquad
F\ =\ f(x)\, dx \wedge dt \ ,\qquad V(\phi)\ =\
\sum_{n\geq 0} a_n e^{2n\rho\phi(x)} \cr
& a_0\ =\ 0\ ,\qquad \epsilon\ \neq\ 0 \ .&\numali\cr}
$$
Moreover, it is convenient to set
$$\eqalignno{
&\alpha\ =\ {4\over\gamma -4}\ ,\qquad \epsilon\ =\ 4-4\rho\ ,\qquad
\rho\ =\ -{2\over \alpha} &\numali \cr
& b_n\ =\ e^{2n\rho\phi_0} ( a_n -\frac12 \delta_{n,2} f_0^2)
\ ,\qquad b_1\ =\ \alpha + \alpha^2\ .\cr}
$$
One then gets
$$\eqalignno{
& g(x)\ =\ 1 - 2m\, x^{\alpha+1} +
\sum_{n\geq 2}{b_n\over (2n-1)\alpha +\alpha^2}\, x^{2(1-n)}\cr
& \phi\ =\ \phi_0 +\frac12 \alpha \log (x) \ ,\qquad
f(x)\ =\ f_0 e^{(2-\epsilon)\phi(x)} &\nameali{solgam}\cr}
$$
where the following conditions must be satisfied: $\alpha+1 <0$ (which
implies $0<\gamma <4$) and $b_n=0$ if $ 2\leq n <n_0$ where $n_0$ is a
fixed number greater than 2 which satisfies $2(1-n_0) < \alpha+1$
(both these conditions come from the requirement that the metric is
asymptotic Minkowskian, i.e. $g(x) \rightarrow 1$ for $x \rightarrow +
\infty$ and that the ({\scaps adm}) mass does not diverge).

Solution 2 is obtained by setting $\phi_0=0$ and $\gamma=2$ (which
implies $\rho=1$ and $\alpha=-2$).
\chapter{Thermodynamics}
On general grounds it must be that $-I = F/T_c = \beta F$ where $F$
is the free energy and $T_c$ the temperature [\Ref{GH}]. From
thermodynamics we
also expect $F= M-T_c S - \sum_i \mu_i {\cal Q}_i$ where $M$ is the
({\scaps adm}) mass, $S$ the entropy, $\mu_i$ the chemical potentials
and ${\cal Q}_i$ the associated charges. Therefore we must find
$$
- I_{\,{\displaystyle\vert}_{\scstyle\rm on\ shell}}\ =\
\beta M - S - \beta \sum_i \mu_i {\cal Q}_i\ .
\efr
We will need to evaluate eq. \shell. A first important observation
is in order. For the flat space solution $g(x) = 1 $ one gets
$$
I_{\,{\displaystyle\vert}_{\scstyle\rm on\ shell}}\ =\ -\int dt dx \,
\del_x \left[ 4 e^{-2\phi} \del_x \phi
\right]_{\,{\displaystyle\vert}_{\scstyle\rm on\ shell}}\ .
\nfr{flatsp}
For $\gamma = 4$ (Solution 1), one has
$$
I_{\,{\displaystyle\vert}_{\scstyle\rm on\ shell}}\ =\ 2\beta Q
e^{-2\phi_0} \left[e^{Qx}\right]^\infty_0\ ;
\efr
for $\gamma=2$ (Solution 2), one has
$$
I_{\,{\displaystyle\vert}_{\scstyle\rm on\ shell}}\ =\ 4\beta
\left[\,x\,\right]^\infty_0\ .
\efr
In both cases the on-shell action diverges, while instead,
 since these solutions describe the flat (empty) space,
one would like to have
$I_{\,{\displaystyle\vert}_{\scstyle\rm on\ shell}} = 0 $.

To get this, we modify the starting action subtracting the flat space
contribution eq. \flatsp. This is exactly what Gibbons and
Hawking were forced to do as well in 4D gravity [\Ref{GH}].
{}From eq. \nshell\ it is indeed obvious that the same divergence
occurs there (in our gauge the {\scaps l.h.s.} of eq. \nshell\ is
exactly eq. \flatsp). Notice that eq. \flatsp\ is a pure boundary term,
and thus does not contribute to the equations of motion.
Therefore, instead of eq. \action, our starting point will be
\noblackbox
$$
I = \int d^2x \sqrt{-g}\, e^{-2\phi} \left[ R +\gamma
(\nabla\phi)^2 -\frac14 e^{\epsilon\phi} F^2 +
V(\phi) + \nabla^a\left( 4 e^{-2\phi}
\nabla_a \phi \right) \right]
\nfr{actthree}
\yesblackbox
for which $I(g=1)_{\,{\displaystyle\vert}_{\scstyle\rm on\ shell}}=0$.
Therefore, instead of eq. \shell, we get
$$
I_{\,{\displaystyle\vert}_{\scstyle\rm on\ shell}}\ =\ -\int dt dx \,
\del_x \left[ 4 e^{-2\phi} (g-1) \del_x \phi + e^{-2\phi} \del_x g
\right]_{\,{\displaystyle\vert}_{\scstyle\rm on\ shell}}
\nfr{newshell}
which, for future convenience, we divide  in two pieces
$$
I_1\ =\ \int dt dx \left[\del_x \left(e^{-2\phi} \del_x g
\right)\right]
\nfr{bound}
and
$$
I_2  =  \int dt dx
\left[ \del_x \left( 4 e^{-2\phi} (g-1) \del_x \phi \right)
\right] \ .
\nfr{bulkshell}

An alternative way to deal with the above discussed infinity is to
leave it there, and interpret it as the divergent contribution to
the chemical potentials associated with a topologically conserved
dilaton charge [\Ref{GP}]. In two dimensions,  the current
$$
J_a\ =\ \epsilon_a^{~b}\, \nabla_b e^{-2\phi(x)}
\efr
is conserved and the associated charge is
$$
D\ =\ \int_\Sigma J_a\, d\Sigma^a
\efr
where $ \Sigma$ is a spacelike hypersurface. The charge
$D=\left[ e^{-2\phi(x)}\right]_{+\infty}$ actually
diverges, since the dilaton is a  a long-range  field, and its
contribution
to the free energy  is explicitly given by eq. \flatsp.
Our prescription of subtracting the divergent term from the
action can be interpreted therefore as eliminating from the free energy
the contribution of the background dilaton field.
This has the advantage that we will deal only with finite quantities
and we will not need to put the theory in a finite
volume [\Ref{GP}].

To explicitly evaluate \newshell, we will work in the euclidean
formalism. We will assume that the time is periodic of period $\beta$
and the space coordinate $x$ takes values between the outer horizon
$x=r_+$ and infinity [\Ref{GH},\Ref{York}]. The horizons are defined
as the roots of the equation $g(x) = 0$ and $r_+$ is the largest one.
For all models we consider here, it is possible to show in the gauge
of eq. \gauge\ that
$$
{4\pi\over \beta}\ = \ \left. {\del g\over\del x}\right\vert_{r_+}\ .
\nfr{eqbeta}

Let's look first to the contribution of the boundary term $I_1$.
Without using any explicit solution but only the definition of $r_+$
and eq. \eqbeta, one finds
$$
I_1\ =\ \beta \left[ e^{-2\phi} \del_x g \right]_{+\infty}
- 4\pi \left[ e^{-2\phi}\right]_{r_+}\ .
\efr
We are then led to the following identifications~\note{This
expression for $M$ has also been obtained in ref. [\Ref{GP}].}
$$
M\ =\ \left[ e^{-2\phi} \del_x g \right]_{+\infty} \ , \qquad\qquad
S\ =\  4\pi \left[ e^{-2\phi}\right]_{r_+}\ .
\efr
We will also show in all the specific cases we will study, that
eq. \bulkshell\ generates the thermodynamic potentials, i.e. $I_2 =
- \beta \sum_i \mu_i {\cal Q}_i$, so that
$$
I_{1\,{\displaystyle\vert}_{\scstyle\rm on\ shell}}\ =\
\beta M - S \ ,\qquad
I_{2\,{\displaystyle\vert}_{\scstyle\rm on\ shell}}\ =\
- \beta \sum_i \mu_i {\cal Q}_i\ .
\nfr{ios}
Finally, we will check that the thermodynamic relations are satisfied.
In particular, when there is only one conserved charge, the electric
charge, one has the following thermodynamical relations
$$
\beta \ =\ \left[ {\del S \over \del M}\right]_{{\cal Q}_{el}} \ ,
\qquad\qquad \mu_{el}\ =\ T_c \left[ {\del S \over \del {\cal Q}_{el}}
\right]_{M}\ .
\nfr{eqtherm}
\section{Solution 1 ($\gamma=4$)}
Let us first apply the previous considerations to the Solution 1 and
in particular to the case of the charged heterotic black-hole:
$$g(x)\ =\ 1 -2m e^{-Qx} + q^2 e^{-2Qx}\ ,\qquad\quad \phi(x)\ =\
\phi_0 - \frac12 Qx\ .
\efr
{}From the previous section one gets
$$\eqalignno{ &e^{Q r_+}\ =\ m+ \sqrt{m^2 - q^2} \ ,\qquad
\beta \ =\ {2\pi \over Q}\,\cdot\, {m+\sqrt{m^2-q^2} \over
\sqrt{m^2 - q^2} } &\nameali{qS}\cr
& M\ =\ 2mQ e^{-2\phi_0}\ ,\qquad\quad
S\ =\ 4\pi e^{-2\phi_0} \left( m + \sqrt{m^2 - q^2} \right)\ .\cr}
$$
The electric charge and chemical potential are given by
[\Ref{YN},\Ref{GP}]
$${\cal Q}_{el}\ =\ \sqrt{2} q Q e^{-2\phi_0}\ =\ {q\over \sqrt{2}
m} M \ ,\qquad
\mu_{el}\ =\ - { \sqrt{2} q\over m + \sqrt{m^2-q^2} }\ .
\efr
One can then check that~\note{In deriving these
results the following identity can be useful:
$m-\sqrt{m^2-q^2}=q^2/(m+\sqrt{m^2-q^2})$.}
$$
I_2\ =\ -\beta\mu_{el} {\cal Q}_{el}\ =\ \beta\cdot 2Q e^{-2\phi_0}
q^2 e^{-Qr_+}\ .
\nfr{qel}
To verify the thermodynamical relations eqs. \eqtherm, one needs to
express $S$ as a function of $M$ and ${\cal Q}_{el}$ as follows
$$
S\ =\ {2\pi\over Q}\left(M+\sqrt{M^2-2{\cal Q}_{el}^2} \right)\ ,
\efr
then it is easy to see that eqs. \eqtherm\ hold.
The above results agree, after the appropriate change of
variables, with those in ref. [\Ref{GP}]~\note{This change of variables
is singular in the limit $m \rightarrow \pm q$. This is the reason why
this limit is unattainable in ref. [\Ref{GP}].}.
Notice that in ref. [\Ref{GP}] these results were obtained by
exploiting thermodynamic
relations rather than our prescription.

Let's consider now the general Solution 1 with
$V(\phi)\ =\ \sum_{n\geq 0} a_n e^{2n\phi(x)}$.  Obviously, it is
not anymore possible to explicitly compute $r_+$ so that the
quantities depending on it will remain in an implicit form. Thus, one
has
$$\eqalignno{ & M\ =\ 2mQ e^{-2\phi_0} &\numali\cr
& S = 4\pi \left[ e^{-2\phi}\right]_{r_+} = 4\pi \left[
e^{-2\phi_0} \left( 2m - q^2 e^{-Qr_+}\right) + \sum_{n\geq 2} { a_n
e^{(n-1)(2\phi_0 - Q r_+)} \over (n-1)Q^2}\right].\cr}
$$
Notice that the ({\scaps adm}) mass doesn't change whereas the entropy
does, as one would expect (for instance, we know that the entropy must
change in the presence of an electromagnetic field, which is the $a_2
\neq 0$ case).
It is easy to check that for
$a_n=0$ ($n\geq 2$), one gets back eq. \qS. Moreover
$$
I_2\ =\ 2\beta Q e^{-2\phi_0} \left(2m-e^{Q r_{+}}\right)\ =\
\beta\,\cdot\, 2Q e^{-2\phi_0} q^2 \left( 1 + d(r_+)\right) e^{-Qr_+}
\efr
where
$$
d(r_+)\ =\ \sum_{n\geq 2} { a_n e^{n(2\phi_0 - Q r_+)}
\over (1-n)q^2 Q^2}\ .
\efr
This formula is very similar to eq. \qel\ which
suggests us to define a renormalized electric charge as follows:
$$\eqalignno{ &q^2_{ren}\ =\ q^2 (1 + d(r_+)) &\numali\cr
& {\cal Q}_{el}^{ren}\ =\ \sqrt{2}\, q_{ren} Q e^{-2\phi_0}\ ,
\qquad\quad \mu_{el}^{ren}\ =\ -\sqrt{2}\, q_{ren} e^{-Qr_+}\cr}
$$
so that, again
$$
I_2\ =\ -\beta\,\mu_{el}^{ren} {\cal Q}_{el}^{ren}\ .
\efr
Finally, one can easily check that
$$
S\ =\ {2\pi\over Q}\left(M+\sqrt{M^2-2({\cal Q}_{el}^{ren})^2}\right)
\nfr{renor}
although it is not anymore possible to check eqs. \eqtherm\ since we
don't know the explicit expression of $r_+$. The formula \renor\
suggests that the loop corrections can be interpreted as
charge renormalization. This is the only reasonable interpretation,
since, aside from the electromagnetic one, there is no other conserved
charge in the game that could give rise to new chemical potentials.
\section{Solution 2 ($\gamma=2$)}
The discussion just done for the Solution 1 can be repeated for the
solutions with $ 0 < \gamma < 4$ and in particular for those with
$\gamma =2$. Although it is possible to give the explicit formul\ae\
for the general case $ 0 < \gamma < 4$ (and the interested reader can
easily work them out), for simplicity here we will just quote those
for the Solution 2 ($\gamma=2$). One has
$$\eqalignno{ & M\ =\ 2m e^{-2\phi_0} &\numali\cr
& S = 4\pi \left[ e^{-2\phi}\right]_{r_+}= 4\pi \left[ e^{-2\phi_0}
\left( 2m r_+ - q^2\right) - \sum_{n\geq 2} {e^{2(n-1)\phi_0} a_n
\over 6 -4n} r_+^{-2n}\right]\cr}
$$
where $ q^2\ =\ \frac14 e^{4\phi_0} f_0^2$. Moreover
$$\eqalignno{ I_2\ &=\ \beta \cdot 4 e^{-2\phi_0} \left( 2m
-r_+\right) &\numali\cr
& =\ \beta \cdot {4q^2 e^{-2\phi_0}\over r_+} \left(1 + {1\over q^2}
\sum_{n\geq2} {a_n e^{2n\phi_0}\over 6 -4n} r_+^{4-2n}\right)\cr}
$$
which suggests to define
$$\eqalignno{ & q^2_{ren}\ =\ q^2 \left(1 + {1\over q^2}
\sum_{n\geq2} {a_n e^{2n\phi_0}\over 6 -4n} r_+^{4-2n}\right)\cr
& {\cal Q}_{el}^{ren}\ =\ 2q_{ren} e^{-2\phi_0}\ ,\qquad\quad
\mu_{el}^{ren}\ =\ -{2q_{ren}\over r_+}&\numali\cr}
$$
so that $ I_2\ =\ -\beta\,\mu_{el}^{ren} {\cal Q}_{el}^{ren}$.
It is also simple to check that
$$
S\ =\ 4\pi  e^{-2\phi_0} (r_+)^2\ =\ 4\pi {e^{-2\phi_0}\over 4}\left(
M + \sqrt{M^2 -({\cal Q}_{el}^{ren})^2 } \right)^2\ .
\efr

Setting $a_n =0$ for $n\geq 2$ and $\phi_0=0$, one gets
$$\eqalignno{ & g(x)\ =\ 1 -{2m\over x} + {q^2 \over x^2}\ ,\qquad
r_+\ =\ m+ \sqrt{m^2 - q^2} &\numali\cr
& \beta \ =\ 2\pi \left(2\sqrt{m^2-q^2} + 2m +  {q^2 \over
\sqrt{m^2 - q^2} } \right) \cr
&M\ =\ 2m\ ,\qquad {\cal Q}_{el}\ =\ 2 q\ ,\qquad \mu_{el}\ =\
- { 2 q\over r_+ } \cr
& S\ =\ 4\pi (r_+)^2 \ =\ \pi \left( M +
\sqrt{M^2 -({\cal Q}_{el})^2 } \right)^2\ .\cr}
$$
These are the well known results for a 4D charged, spherically
symmetric black-hole, {\it i.e.} the Reissner-Nordstr{\o}m black-hole,
obtained now from a two dimensional analysis.
\acknowledgements
We would like to thank E.~Witten for some useful comments.
C.N. is partially supported by the Ambrose Monell Foundation.
The research of A.P. is supported by an INFN fellowship and partially
by the NSF grant PHY90-21984.
\references
\beginref
\Rref{Witten}{E.~Witten, ``{\sl String Theory and Black-Holes\/}",
Phys. Rev. {\bf D44} (1991) 314.}
\Rref{LN}{O.~Lechtenfeld and C.~Nappi, ``{\sl Dilaton Gravity and
No-Hair Theorem in Two Dimensions\/}", preprint IASSNS-HEP-92/22,
hep-th/9204026, March 1992, to appear in Phys. Lett. {\bf B}.}
\Rref{YN}{M.~McGuigan, C.~Nappi and S.~Yost, ``{\sl Charged
Black-Holes in Two-Dimensional String Theory\/}", Nucl. Phys.
{\bf B375} (1992) 421.}
\Rref{GP}{G.W.~Gibbons and M.J.~Perry, ``{\sl The Physics of 2d
Stringy Space-Time\/}", preprint hep-th/9204090, March 1992.}
\Rref{GH}{G.W.~Gibbons and S.W.~Hawking, ``{\sl Action Integrals and
Partition Functions in Quantum Gravity\/}", Phys. Rev. {\bf D15} (1977)
2752.}
\Rref{York}{B.F. Whiting and J.W. York, ``{\sl Action Principle and
Partition Function for the Gravitational Field in Black-Hole
Topologies\/}", Phys. Rev. Lett. {\bf 61} (1988) 1336.}
\Rref{Teit}{M.~Banados, C.~Teitelboim and J.~Zanelli, ``{\sl The
Black-Hole in Three Dimensional Space-Time\/}", preprint
IASSNS-HEP-92/29, hep-th/9204099, May 1992.}
\Rref{Wilc}{F.~Wilczek, lecture notes at Princeton University, 1992,
unpublished.}
\Rref{Jactei}{R.~Jackiw, in {\sl Quantum Theory of Gravity\/}, ed.
S.~Christensen, Hilger, Bristol, UK, 1984;\newline
C.~Teitelboim, ``{\sl Gravitation and Hamiltonian Structure in Two
Space-Time Dimensions\/}", Phys. Lett. {\bf B126} (1983) 41;\newline
R.B. Mann, ``{\sl Lower Dimensional Gravity\/}", preprint
WATPHYS-TH-91-07, 1991.}
\Rref{Callan}{C.G. Callan, S.B. Giddins, J.A. Harvey and S. Strominger,
``{\sl Evanescent Black-Holes\/}", Phys. Rev. {\bf D45} (1992) 1005;
\newline L. Susskind and L. Thorlacius, ``{\sl Hawking Radiation and
Back-Reaction\/}", preprint SU-ITP-92-12, hep-th/9203054, March 1992;
\newline  S.P. De Alwis, ``{\sl Black Hole Physics from Liouville
Theory\/}", preprint COLO-HEP-284, hep-th/9206020, June 1992.}
\Rref{Odin}{J.G.~Russo and A.A.~Tseytlin, ``{\sl Scalar-Tensor
Quantum Gravity in Two Dimensions\/}", preprint
SU-ITP-92-2, DAMTP-1-1992, hep-th/9201021, January 1992; \newline
T.T. Burwick and A.H. Chamseddine, ``{\sl Classical and Quantum
Considerations of Two Dimensional Gravity\/}", preprint
ZU-TH-4/92, hep-th/9204002, March 1992; \newline
E.~Elizalde and S.D.~Odintsov, ``{\sl One-Loop
Divergences in Two-Dimensional Maxwell-Dilaton Quantum Gravity\/}",
preprint UB-ECM-PF 92/17, June 1992.}

\endref
\ciao
